\documentclass[twocolumn]{revtex4}
\usepackage{times}
\usepackage{color}
\usepackage{amsmath}
\usepackage{amssymb}
\usepackage{graphicx}

\usepackage{color}

\newcommand{\be}{\begin{equation}}
\newcommand{\ee}{\end{equation}}
\newcommand{\bea}{\begin{eqnarray}}
\newcommand{\eea}{\end{eqnarray}}
\newcommand{\bc}{\begin{center}}
\newcommand{\ec}{\end{center}}

\begin{document}
\title{Are priors responsible for cosmology favoring additional neutrino species?}
\author{Alma X. Gonz\'alez-Morales$^{1}$, Robert Poltis$^{2}$, Blake D. Sherwin$^{3}$, Licia Verde$^{4}$}
\affiliation{$^1$ Instituto de Ciencias Nucleares, Universidad Nacional Aut\'onoma de M\'exico, Apdo. 70-543, CU, 04510
M\'exico D.F. Part of the Collaboration Instituto Avanzado de Cosmolog\'ia.\\
$^2$ HEPCOS, Department of Physics,
SUNY at Buffalo, Buffalo, NY 14260-1500, USA.\\
$^3$ Department of Physics, Jadwin Hall, Princeton University, Princeton, NJ 08544, USA.\\
$^{4}$ ICREA \& ICC, University of Barcelona (IEEC-UB), Marti i Franques 1, Barcelona 08028, Spain.}

\date{\today}
\begin{abstract}
It has been suggested that both recent cosmological data and the results of flavor oscillation experiments (MiniBooNE and
LSND) lend support to the existence of low-mass sterile neutrinos. The cosmological data appear to weakly favor additional forms of
radiation in the Universe, beyond photons and three standard neutrino families. We reconsider the cosmological
evidence by making the resulting confidence intervals on the additional effective neutrino species  as prior-independent as
possible. We find that, once the prior-dependence is removed, the latest cosmological data show no evidence for
deviations from the standard  number of neutrino species. 
\end{abstract}
\pacs{?}
\maketitle
\section{Introduction}
\label{Intro}
The standard model for particle physics has three massless neutrinos. Beyond the standard model physics is needed
to give neutrinos a non-zero mass and hence explain measurements of neutrino oscillations. It is therefore reasonable, when considering extensions of the standard model, to explore the possibility of more than three neutrinos species.
A deviation from the standard number of neutrino species $N_{\nu}=3$ affects the expansion history of the early universe, which in turn modifies nucleosythesis constraints. Observations of the Cosmic Microwave Background (CMB) and of large-scale structure also probe
the radiation density at matter-radiation equality and at recombination; this physical energy density in
relativistic particles can be  expressed in terms of  the energy density in photons (which is highly constrained by the
measurement of the CMB temperature) and the effective number of neutrino species $N_{\rm eff}$ which, even in the
standard scenario \footnote{In the standard scenario   $N_{\nu}=3$ and  $N_{\rm eff}=3.04$.},  differs from the number of
neutrino species $N_{\nu}$  to account for QED effects and for neutrinos being not completely 
decoupled during electron-positron annihilation (see \cite{lesgourguespastor} for a review and refs. therein). Any light
particle that does not couple to electrons, ions and photons, such as a sterile neutrino, will act as an 
additional relativistic species. Deviations from the standard $N_{\rm eff}$ value  can  also arise from decay of dark
matter particles,  quintessence, light axions or other exotic models \cite{exotic}.
Nucleosynthesis  and CMB constraints on $N_{\nu}$ rely on completely different physics and correspond to very different
epochs in the Universe's evolution, providing an overall consistency check.

Recently, there has been renewed interest in the deviations for the standard  number of  neutrino species, see e.g.,
\cite{nature}. Extra neutrinos could explain for example the LSND results and some of the latest MiniBoone results
\cite{miniboone, miniboone2,giunti}. The anomalous X-ray narrow emission feature  from the dwarf spheroidal galaxy
Willman 1 has also been explained in terms of sterile neutrinos \cite{willman1}.

The current situation is summarized in Ref. \cite{Hamann/etal:2010}. The authors consider the latest cosmological constraints
on the number of neutrino species and their possible connection with  particle physics experiments. They find that
recent cosmological data, alone or in combination with big-bang nucleosynthesis constraints, weakly favor extra radiation in the
Universe beyond photons and ordinary neutrinos, lending support to the existence of low-mass sterile neutrinos.  They find that the amount of additional radiation corresponds to one
extra $N_{\rm eff}$. An extra species of low mass sterile neutrinos would fit both cosmological data and flavor
oscillation results. 

Note that in some analyses of cosmological data the standard $N_{\rm eff}$ value is ruled out at better than 95\%
confidence, e.g. \cite{concha}, while in other analyses it is not, e.g. \cite{reid}.
More recently, the latest CMB  temperature data, including high-resolution measurements of the damping tail,  were combined in
Ref.~\cite{Hou/etal:2011} to show that this preference for extra neutrino species in the cosmological data comes from
their contribution to the expansion rate prior to recombination. This analysis also rules out the standard neutrino
number,
$N_{\rm eff}$, at the $95 \%$ and $98.4 \%$ confidence level with data sets consisting of CMB data  and of a combination of CMB, Large-scale structure
and the Hubble constant  respectively. The physical interpretation offered is that a larger value of $N_{\rm eff}$ produces an increased Silk
damping at high $\ell$ and  this seems to improve the fit to the data.

Given the importance of the consequences of a deviation from the standard effective number of neutrino  species and the 
reported hints for extra radiation from  precision cosmological data, we reconsider the cosmological constraints on  $N_{\rm eff}$.
\section{Method}
Early work (\cite{previous, debernardis, reid, hannestad, wmap, concha} and references therein)  as well as more recent papers \cite{Hou/etal:2011,DunkeleyACT} have reported
constraints derived in the Bayesian framework. In this framework constraints are obtained from the posterior
distribution, which is marginalized over the uninteresting parameters to derive  the desired confidence intervals. While
this approach is extremely powerful and enables one to do statistical inference, the posterior distribution  depends on the priors chosen for each of the parameters, and this choice is often arbitrary. Furthermore, when deriving
confidence intervals, the marginalization procedure weights
parameters  by the posterior volume, rather than just by how well  the parameters fit the data.
 When considering a high-dimensional  parameter space, especially if the data-sets are not highly constraining, the 
reported confidence regions on a parameter can depend strongly on the prior choice and on  the resulting prior volume effects.

Moreover the widely used software ``Get Dist" -- part of the CosmoMC package \cite{cosmomc}-- provides  by default  the
Bayesian central credible interval while the minimal credible interval may be more suitable for statistical inference.
As noted in \cite{hamann1} these two confidence intervals differ significantly if the posterior is skewed and the effect
is large especially for $N_{\rm eff}$ if the data-sets chosen are not very constraining. As more data sets are
considered, degeneracies are reduced, the posterior become  nearly Gaussian and the two confidence intervals become
closer.

If a result of cosmological analysis, such as any evidence for $N_{\rm eff} > 3.04$, turns out to be  driven by any of
the effects above, it should be interpreted with  extra care.
 
The dependence on the (arbitrary) choice for the prior, on the posterior volume effect in marginalization and the
ambiguity of central vs minimal   credible interval can be removed by making the analysis as prior-independent as
possible.

In the frequentist approach the generalized likelihood ratio is widely used  to report confidence intervals. The
likelihood ratio is a well established technique to compute frequentist confidence interval when only one parameter is
considered, but it lacks  of  a feature equivalent  to Bayesian marginalization when dealing with  high-dimensional
(multi-parameters) problems. The generalized (or profile) likelihood ratio on the other hand includes such a feature.
 
Fortunately, it is possible to extract profile likelihood ratio-based confidence intervals  from  the standard  output of cosmological  analyses implemented with Markov Chain Monte Carlo methods \cite{reid}.

To perform our  Markov Chain Monte Carlo analysis, we use the publicly available CosmoMC package \cite{cosmomc}. The output from a CosmoMC analysis for a given data set contains the likelihood ($L$) value for each accepted set of parameters. Since we are intersted only in constraints of $N_{\rm eff}$, we find the maximum likelihood ($L_{N_{\rm eff}}$) value as a function of $N_{\rm eff}$ over the parameter space sampled in the chain, without regard to the values that the rest
of the parameters have. To perform this maximization in practice we bin in $N_{\rm eff}$ with a bin width of $0.5$. We verified that the results do not depend  strongly on the choice of binning (as long as bins are not so small that they give a very noisy estimate and not so large that they erase any signal).
We obtain the profile likelihood ratio by considering
 $\ln\left(L_{N_{\rm eff}}\,/\,L_{\rm max}\right)$ as a function of $N_{\rm eff}$; where $L_{\rm max}$ is the maximum likelihood in
the entire chain. 

 Then, always following \cite{reid},  
  we use the pseudo-chisquare defined as $\ln\left(L_{N_{\rm eff}}\,/\,L_{\rm max}\right)=1/2 \, \chi^2$, so that $\Delta
\ln\left(L_{N_{\rm eff}}\,/\,L_{\rm max}\right)= 0.5 $  and $\Delta \ln\left(L_{N_{\rm eff}}\,/\,L_{\rm max}\right)= 2 $ correspond to
the $68.3 \%$ and $95.4\%$ confidence regions respectively. Finally we report our results in terms of the probability
for $\Delta N_{\rm eff} = N_{\rm eff}-3.04$  for each data set considered. The correspondence between $\chi^2$ intervals and probability is done -as it is customary in this context- assuming Gaussian  statistics; the confidence intervals we report  therefore could be slightly underestimated  because of our Gaussian assumption.

We consider the following data-set combinations: Wilkinson Microwave Anisotropy  Probe 7-year data
\cite{clwmap7,Komatsuwmap7} (WMAP) in combination with a Hubble constant prior of \cite{h0},   WMAP in combination with
the baryon acoustic oscillation constraints from the Sloan Digital Sky Survey data release 7 (SDSS DR7) luminous red galaxies (LRG)
\cite{bao} (WMAP+BAO), WMAP in combination with the SDSS DR7 LRG halo
power spectrum of Ref.\cite{reidLRG} (WMAP+LRG), and the luminosity distance measurements of type 1A supernovae of Ref.
\cite{sn} (SN).
We also consider  high resolution CMB experiments for which data, angular power spectra in the multipole range of
interest and likelihood functions were released publicly at the time of writing: ACBAR\cite{ACBAR} and ACT
\cite{DunkeleyACT}.  
In our analysis we keep the Helium mass fraction $(Y_P$) fixed to the concordance value. Some analyses
\cite{DunkeleyACT, Hou/etal:2011} allow $Y_P$ to vary. Adding yet another free parameter in the analysis opens up
extra degeneracies (in particular $N_{\rm eff}$ and $Y_P$ are anti-correlated), amplifying and  possibly exacerbating
the prior effects  that, we suspect, may ``bias'' the marginalized posteriors.

In our chains we use a flat prior on $H_0$ rather than a flat prior on the angular diameter distance to the last scattering surface  as in the standard CosmoMC implementation. In fact the fitting formula used by the code to compute the  angular diameter distance is accurate for standard number of neutrino species but it is not otherwise.  In practice, using a flat prior on  this quantity is equivalent to using a non-flat and somewhat complex prior on the true angular diameter distance for non-standard number of neutrinos.  This effect goes away when using a flat prior on $H_0$ and bypassing this analytic fit to the angular diameter distance (see e.g., \cite{debernardis} for discussion on this). For our specific application, using the profile likelihood the choice of prior does not matter and indeed we have verified that by running chains with both a flat prior on $H_0$ and on  the angular diameter distance for some selected  cases: the profile likelihood ratio results are indistinguishable, while the marginalized posteriors are not. 
\section{Results and conclusions}
\begin{figure}
 \begin{center}
    \includegraphics[height=2.1in,width=2.9in]{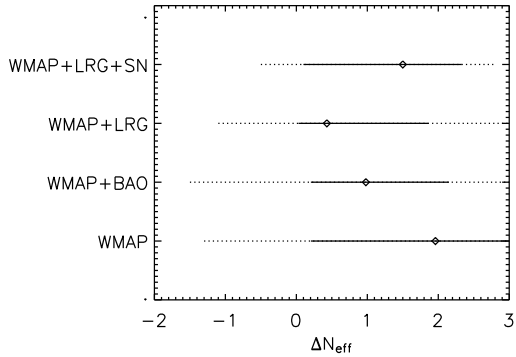}
    \caption{One and two $\sigma$ constraints on the number of additional neutrino species from several datasets including combinations of WMAP7, LRG, BAO and SN. Note that the value corresponding to no extra species, $\Delta N_{\rm eff} =0$, is always well within the 2-$\sigma$ interval and always very close to the 1-$\sigma$ interval.}
    \label{fig:summarywmap}
  \end{center}
\end{figure}

\begin{figure}
 \begin{center}
    \includegraphics[height=2.3in,width=3in]{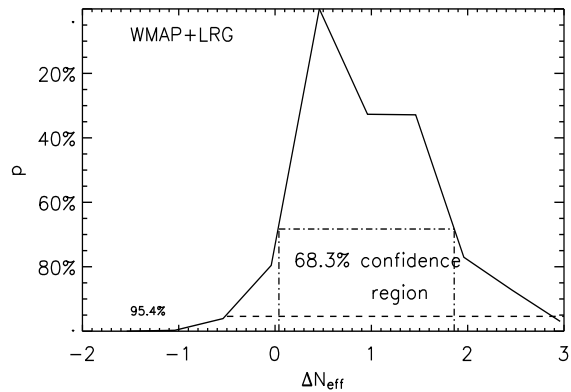}
    \caption{As an illustrative example we show the probability curve obtained from the profile likelihood ratio for the data combination  WMAP+LRG. The curve is jagged because of the binning procedure used to compute the profile likelihood. We have checked that changing the bin size (within reason) only slightly changes the shape of the curve and does not change the inferred  confidence  regions.}
    \label{fig:WMAP-LRG}
  \end{center}
\end{figure}

We begin by considering WMAP7 CMB data, both alone  and in combination with large-scale structure and supernova datasets.
A compilation of the constraints from the different datasets we considered is shown in figure
\ref{fig:summarywmap}.  Note that for all datasets, although the maximum likelihood value is always at
$\Delta N_{\rm eff}>0$,   $\Delta N_{\rm eff}=0$ is always well within the  95\% error bars. All these combinations have
the WMAP7 data in common are are hence not at all statistically independent, so it is not surprising   that if in the WMAP data alone the maximum likelihood value is
$\Delta N_{\rm eff}>0$, this persists in all  combinations. As an illustrative example on how these error-bars were
derived from the profile likelihood ratio,  in Fig. \ref{fig:WMAP-LRG}  we show the probability curve obtained from the
profile likelihood ratio for the combination WMAP7+LRG. This can be compared directly with the results of \cite{reid}.  
\begin{figure}
 \begin{center}
    \includegraphics[height=2.3in,width=3in]{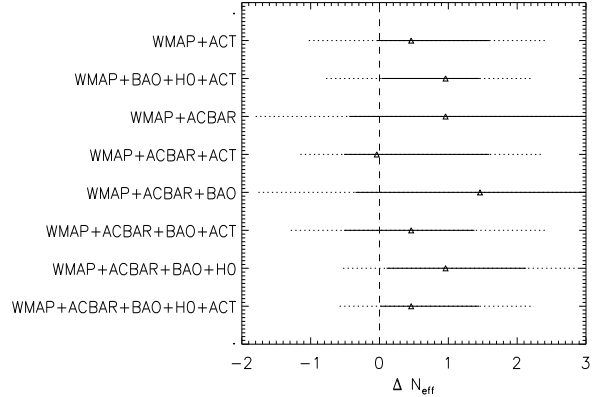}
    \caption{Overview of one and two $\sigma$ constraints on extra number of neutrino species from several dataset combinations involving WMAP7 in combination with higher resolution CMB experiments.  Note that the point corresponding to no additional relativistic species, $\Delta N_{\rm eff} =0$, is always well within the 2-$\sigma$ interval and in few cases  within the 1-$\sigma$ interval.  the fact that all constraints seems to be systematically shifted (although by less than 1-$\sigma$) towards $\Delta N_{\rm eff} >0$ is due to the fact that these are {\it not} independent constraints: the WMAP data, which has a lot of statistical power, is common to all combinations of datasets.}
    \label{fig:summaryallcmb}
  \end{center}
\end{figure}
\begin{figure}
 \begin{center}
    \includegraphics[height=2.2in,width=3in]{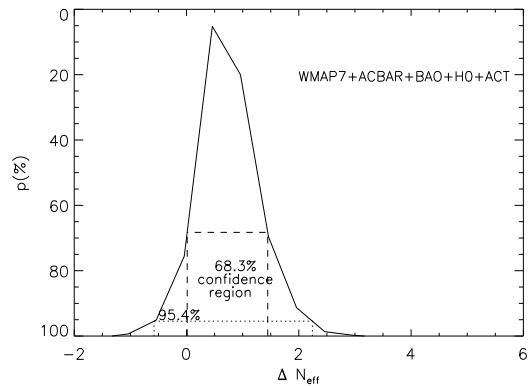}
    \caption{As an illustrative example we show the probability curve obtained from the profile likelihood ratio for  one of the data combinations. }
    \label{fig:allcmb}
  \end{center}
  \end{figure}
We then consider the constraints from data-sets combination involving WMAP7 and higher-resolution CMB experiments. The constraints from the different dataset combinations are summarized in Fig. \ref{fig:summaryallcmb} and an illustrative example of probability curve for one combination is shown in Fig. \ref{fig:allcmb}.
Again, the standard value $\Delta N_{\rm eff}=0$ is always well within the 95.4\% confidence region and in many cases within the 1-$\sigma$ region.

If we had looked at the marginalized posterior rather than the profile likelihood, the following data-set combination would have given an almost  2-$\sigma$ indication of $\Delta N_{\rm eff}>0$: WMAP7, WMAP7+BAO+H0, WMAP7+LRG+H0 (see \cite{wmap}). Recall also that looking at the marginalized posterior, for the combinations
CMB+H0+SN+BAO and CMB+H0+SN+LSS, \cite{concha} find a 2-$\sigma$ signal for $\Delta N_{\rm eff}>0$ and that \cite{Hou/etal:2011} find  a  2-$\sigma$ signal for $\Delta N_{\rm eff}>0$ for WMAP7+ ACT+ACBAR+BAO+H0, although using different priors and slightly different parameter sets (i.e. the treatment of $Y_P$).

We can also try to compare with \cite{Hamann/etal:2010} and a more recent work \cite{olga}. In these references the authors also allow the neutrino masses to vary. Current cosmological data do not  show any evidence for a non-zero neutrino mass and do not have enough statistical power  to detect a sum of neutrino masses ($\Sigma$) below $0.3$ eV (e.g., \cite{reid, lahav}). Including both $N_{\rm eff}$ and the sum of neutrino masses as free parameters would add new degeneracies and exacerbate further the prior volume effect in the marginalized posterior distributions, especially since $N_{\rm eff}$ and $\Sigma$ are positively correlated.  To compare with our analysis  we need to consider their constraints for $\Sigma=0$. They find still almost 2-$\sigma$ indication of  $\Delta N_{\rm eff}>0$ \footnote{Recall that the 1-$\sigma$ joint contour in 2 dimensions is much  closer to the 2-$\sigma$ marginalized in 1-dimension than to the 1-$\sigma$ one.}. Our analysis seems to indicate that this evidence may be driven by prior volume effects.

To summarize, the effective number of neutrino species parameterizes any non-standard early-Universe expansion rate. Evidence for $\Delta N_{\rm eff} \neq 0$ could be interpreted as extra (sterile) neutrino species but also as  any light particle that does not couple to electrons, the decay of dark matter particles, early quintessence and other phenomena. 
Recent cosmological analyses carried out in the Bayesian framework have reported hints of $\Delta N_{\rm eff} >0$; extra neutrino species  could explain  some recent claims by particle physics experiments (LSND, MiniBoone). 
In the Bayesian framework, when considering a high-dimensional parameter space as in this case, especially if the data-sets are not highly constraining and cosmological degeneracies are present,  the reported  marginalized confidence regions on a parameter can depend strongly on the prior choice and on prior volume effects on other parameters.
If a result of cosmological analysis, such as evidence for $\Delta N_{\rm eff}>0$,   turns out to be driven by any of these effects,  it should be interpreted with extra care.
We have presented a way to make the cosmological analysis as prior-independent as possible; we borrowed  from the frequentist approach the so-called generalized likelihood ratio to report confidence intervals.
We have considered a suite of  cosmological data sets and data sets combinations and found that prior-independent confidence intervals for $\Delta N_{\rm eff}$ do not show any evidence of additional effective neutrino species.  Our findings seems to indicate that any evidence for  $\Delta N_{\rm eff}>0$ may be driven by prior effects. As better data become available, the likelihood should overcome the prior and the posterior should become nearly Gaussian, so that this effect should gradually disappear. In the meantime we advocate the use of the generalized likelihood ratio as a useful check of how dependent cosmological results are on the choice of priors.
\acknowledgments{\small This work started as a tutorial exercise for the ``Statistical and numerical tools in cosmology"
lecture course \footnote{Lecture notes are available at http://bccp.lbl.gov/beach\_program/presentations11.html},
at the ``Essential Cosmology for the next generation" 2011 school   in Puerto
Vallarta, M\'exico. We thank the organizers, the Berkeley Center for Cosmological Physics and the 
Instituto Avanzado de Cosmolog\'ia, for
a successful school and  for  creating such a stimulating environment.  AXGM is supported by DGAPA-UNAM IN115311
grant and CONACyT scholarship. BDS is supported by a National Science Foundation Graduate Research Fellowship. LV is supported by FP7-IDEAS-Phys.LSS 240117 and MICINN grant AYA2008-03531. We acknowledge the use of the Legacy Archive for Microwave Background Data Analysis (LAMBDA). Support
for LAMBDA is provided by the NASA Office of Space Science.}
\addcontentsline{toc}{section}{References}
\bibliographystyle{JHEP}

\begin{thebibliography}{200}
\bibitem{lesgourguespastor} Lesgourgues, J. Pastor S.,  2006, Phys.Rep. 429, 307
\bibitem{exotic}
Bonometto S., Pierpaoli E., 1998, New Astronomy, 3 , 391; Lopez R.E.,Dodelson S., Sherrer R.J., Turner M.S., 1998, Phys.
Rev. Lett. 81,3075; Hannestad S., 1998, Phys.Rev.Lett. 80, 4621 ; Kaplinghat M.,  Turner M.S., 2001, Phys. Rev. Lett.
86, 385 ;   H.~Davoudiasl, arXiv:0705.3636
\bibitem
{nature} Hand, E., Nature, 2010, 464, 334
\bibitem
{miniboone} The 
MiniBooNE Collaboration 2010, PRL, 105.181801
\bibitem
{miniboone2} Gninenko, S.~N.\ 2011, PRD, 
83, 015015 ; Akhmedov, E., \& Schwetz, T.\ 2010, Journal of High Energy Physics, 10, 115 
\bibitem
{giunti} Giunti, C., Leveder, M., 2010,  PRD, 82,053005 
\bibitem
{willman1} Loewenstein, M., \& Kusenko, A.\ 2010, ApJ, 714, 652 
\bibitem
{Hamann/etal:2010} Hamann, J., Hannestad, 
S., Raffelt, G.~G., Tamborra, I., 
\& Wong, Y.~Y.~Y.\ 2010, Physical Review Letters, 105, 181301 
\bibitem{concha} Gonzalez-Garcia M.~C.,  Maltoni M.,  Salvado J., JHEP, 1008:117, 2010
\bibitem
{reid} Reid, B.~A., Verde, L., 
Jimenez, R., \& Mena, O.,  2010, JCAP, 1, 3 
\bibitem{Hou/etal:2011} Hou, Z., Keisler, R., Knox, L., Millea, M., \& Reichardt, C.\ 2011, arXiv:1104.2333 
\bibitem{DunkeleyACT} Dunkley et al., 2011, arXiv:1009.0866

\bibitem{previous}
Hannestad, S.\ 2003, Journal 
of Cosmology and Astro-Particle Physics, 5, 4;  S.~H.~Hansen, G.~Mangano, A.~Melchiorri, G.~Miele and O.~Pisanti, 
Phys.\ Rev.\  D {\bf 65} (2002) 023511; Hannestad, S.\ 2005, Journal of Cosmology and Astro-Particle Physics, 2, 11 ; 
E.~Pierpaoli, Mon. Not. Roy. Astron. Soc., 342, L63; K. Ichikawa, M. Kawasaki and F. Takahashi,  JCAP 0705, 007 (2007);
Crotty, P., Lesgourgues, 
J., \& Pastor, S.\ 2003, \prd, 67, 123005; Elgar{\o}y, {\O}., \& Lahav, O.\ 2003, Journal of Cosmology and
Astro-Particle Physics, 4, 4

\bibitem
{debernardis}De Bernardis, F., 
Melchiorri, A., Verde, L., \& Jimenez, R., 2008, JCAP, 3, 20 

\bibitem
{hannestad} Hamann, J., Hannestad, 
S., Lesgourgues, J., Rampf, C., \& Wong, Y.~Y.~Y.\ 2010, JCAP, 7, 22 



\bibitem
{wmap} Komatsu, E., et al.\ 
2011, ApJS, 192, 18; 

\bibitem
{cosmomc} Lewis, A., \& Bridle, S.\ 2002, PRD, 66, 103511 
\bibitem
{hamann1} Hamann, J., Hannestad, 
S., Raffelt, G.~G., \& Wong, Y.~Y.~Y.\ 2007, JCAP, 8, 21 

\bibitem{Komatsuwmap7} Komatsu, E., et al.\ 2011, ApJS, 192, 18 
\bibitem{clwmap7}Larson, D., et al. 2011, ApJS, 192, 16 
\bibitem{h0} Riess, A. G., Macri, L., Casertano, S., Sosey, M., Lampeitl, H., 
Ferguson, H. C., Filippenko, A. V., Jha, S. W., Li, W., 
Chornock, R., \& Sarkar, D. 2009, ApJ, 699, 539 
\bibitem{bao} Percival, W.~J., et  al.\ 2010, MNRAS, 401, 2148 
\bibitem{reidLRG} Reid, B.~A., et al.\ 2010, MNRAS, 404, 60 
\bibitem{sn} Kowalski, M., et al.  2008, ApJ, 686, 749; Hicken et al. 2009, ApJ, 700, 1097 
\bibitem{ACBAR} Reichardt et al. 2009, ApJ,  694, 1200

\bibitem{olga} Giusarma E., Corsi,M.,  Archidiacono, M.,  de Putter, R.,  Melchiorri, A.,  Mena, O.,  Pandolfi, S., arxiv:1102.4774 
\bibitem{lahav} Swanson, M.~E.~C., Percival, W.~J., \& Lahav, O.\ 2010, MNRAS, 409, 1100 


\end{thebibliography}

\end{document}